\documentstyle[aps,prb,preprint,amsfonts,tighten]{revtex}
\def\d{{\rm d}}
\def\i{\ifmmode{\rm i}\else\char"10\fi}
\def\e{{\rm e}}
\begin{document}
%
%
\draft
\title{The functional--analytic versus the functional--integral\\
approach to quantum Hamiltonians:\\
The one--dimensional hydrogen atom\footnote{Dedicated to the memory of
Josef Meixner (24 April 1908 --- 19 March 1994).}
}
\author{Werner Fischer, Hajo Leschke and Peter M\"uller}
\address{Institut f\"ur Theoretische Physik,
Universit\"at Erlangen--N\"urnberg,\\
Staudtstra{\ss}e 7, D--91058 Erlangen, Germany}
\date{23 January 1995}
\maketitle
\begin{abstract}
The capabilities of the functional--analytic and of the
functional--integral approach for the construction of the Hamiltonian
as a self--adjoint operator on Hilbert space are compared
in the context of non--relativistic quantum mechanics.
Differences are worked out by taking the
one--dimensional hydrogen atom as an example, that is, a point mass
on the Euclidean line subjected to the inverse--distance potential. This
particular choice is made with the intent to clarify a long--lasting
discussion about its spectral properties. In fact, for the four--parameter
family of possible Hamiltonians the corresponding
energy--dependent Green functions are derived in closed form. The
multiplicity of Hamiltonians should be kept in mind when modelling certain
experimental situations as, for instance, in quantum wires.
\end{abstract}
%
%
\section{Introduction}

This contribution is about a basic aspect of
non--relativistic quantum mechanics of a
point particle in $d$--dimensional Euclidean space $ {\Bbb{R}}^{d} $
subjected to a real--valued potential $ V(q) $. The first task in
characterizing the associated quantum system is to define
its Hamiltonian $ H $ as a self--adjoint operator acting on the Hilbert
space $ L^{2}({\Bbb{R}}^{d}) $ of complex--valued functions which are
square--integrable with respect to the Lebesgue measure $ \d^{d} q $ in
$ {\Bbb{R}}^{d} $. Principally, there are two different approaches to do so.
The functional--analytic approach, see e.g.\ Refs.\
\onlinecite{AkhiezerG,ReedSII}, is concerned
with constructing $ H $ as a self--adjoint extension of Schr\"odinger's formal
differential
operator $ (-1/2)\Delta + V(q) $. Here $ \Delta $ denotes the Laplacian
in $ {\Bbb{R}}^{d} $ and units are chosen such
that the mass of the particle and Planck's constant $ \hbar $ are equal
to unity. In the functional--integral approach, see e.g.\ Refs.\
\onlinecite{EzawaK,McKean77,Carmona79,CarreauF90a},
the operator $ H $ is defined as the (negative) generator of a
self--adjoint strongly continuous one--parameter semigroup on
$ L^{2}({\Bbb{R}}^{d}) $ which, in turn, is defined as an average of a
suitable functional over the paths $ x: t \mapsto x(t) $ of a suitable
Markov process in $ {\Bbb{R}}^{d} $ indexed by time $ t\ge 0 $.
Under rather mild assumptions on the
potential $ V $, both approaches yield naturally the same unique
Hamiltonian $ H $. For example, this is the case if $ V $ is Kat\^o
decomposable, that is, if $ |V| - V $ and $ |V| + V $ are in the
Kat\^o and local Kat\^o class over $ {\Bbb{R}}^{d} $,
respectively.\cite{Simon82}
The equivalence of both approaches is then concisely expressed by
the standard Feynman--Kac formula
\begin{equation} \label{feynmankac}
\langle {  q}_{b}|\exp \left\{ -t\, H\right\} |{  q}_{a} \rangle
= \hspace{-2mm}\int\limits_{{\cal C}({\Bbb{R}}^{d},{  q}_{a})}\hspace{-4mm}
\d W^{d}[{  x}]\;\delta^{(d)} ({  x}(t)-{  q}_{b})\;
\exp \left\{-\int\limits_{0}^{t}\d t'\;V({  x}(t'))\right\} \; .
\end{equation}
Here we have used Dirac's notation for the integral kernel of the
semigroup $ \{ \e^{-tH}\}_{t\ge 0} $.
Moreover, $ \d W^{d} $ denotes Wiener's probability measure, which is
concentrated on the set $ {\cal C}({\Bbb{R}}^{d} , q_{a}) $ of continuous paths
$ x $ in $ {\Bbb{R}}^{d} $ starting from $ q_{a} \in {\Bbb{R}}^{d} $, that is,
$ x(0)=q_{a} $.
The Dirac delta function accounts for the fact that the paths are conditioned
to arrive at $ q_{b} \in {\Bbb{R}}^{d} $ at time $ t $. In other words,
the path average is effectively taken with respect to the Brownian
bridge.\cite{Simon79,Roepstorff,Protter}

As a prominent example we consider the inverse--distance potential
$ V(q)=-\gamma /|q|,\;\gamma \in{\Bbb{R}} $. Even though the
coupling constant $ \gamma $ may take on both signs and
this potential coincides
only for $ d=3 $ with the Newton--Coulomb potential on $ {\Bbb{R}}^{d} $, we
will refer to the associated quantum system as the $d$--dimensional
hydrogen atom for all $ d\ge 1 $. In what follows, an important point
is that $ -\gamma /|q| $ is Kat\^o decomposable if and only if $ d\ge 2 $.

The answer to the question, what happens in $ d=1 $, will be our main
concern here. In fact, a study of the one--dimensional hydrogen atom is of
interest for at least two reasons. First, it nicely illustrates
subtleties in the functional--analytic and in the functional--integral
approach, which are also
present in other systems with a sufficiently singular potential.
These subtleties, while technical in nature, have important physical
consequences. Only if the well--established mathematical techniques for
handling them are generally accepted, one may hope for an end of a
long--lasting discussion.\cite{Loudon,Andrews66,HainesR,Andrews76,%
GomesZ,Andrews81,Moss87,Davtyan,HammerW,Andrews88,NunezV88,BoyaK,%
NunezV89,OsegueraL,Moshinsky93,Newton,Moshinsky94}
Second, the one--dimensional hydrogen atom is of physical relevance in
relation with atoms or excitons in the presence of strong magnetic
fields\cite{ElliottL,RuderW} or in quantum
wires.\cite{LeeS,Bryant,BanyaiG,Abe,OgawaT}
It is also discussed in the semiclassical approximation of
quantum dynamics\cite{SuarezN} and as a model for an electron near the
surface of liquid helium.\cite{SavilleG}

We will see that the standard functional--analytic approach provides a
four--parameter family $ \{ H_{\bf B} \} $ of self--adjoint
extensions of the formal differential operator
$ (-1/2)\d^{2}/\d q^{2} -\gamma /|q| $. As is suggested from the
singularity of the potential, different $ H_{\bf B} $ simply
correspond to different boundary conditions to be imposed on the wave
functions $ \varphi \in L^{2}({\Bbb{R}} ) $ at the origin of the
real line $ {\Bbb{R}} $. In contrast, our functional--integral approach, which
seeks to keep the {\sc rhs} of (\ref{feynmankac}) finite by excluding
appropriate paths, will yield only one element of the family
$ \{ H_{\bf B} \} $, namely $ H_{\rm D} $, the one corresponding to a
Dirichlet condition at $ q=0 $. Finally, we take the opportunity
to rederive the early closed--form expression\cite{Meixner} for the
energy--dependent Green function $ \langle q_{b}| (H_{\rm D} -E)^{-1}
|q_{a} \rangle $ by a genuine functional--integral calculation.

In order to avoid a possible confusion, we remark that the divergency
of the {\sc rhs} of (\ref{feynmankac}) for the attractive case $ \gamma >0 $
of the potential $ V(q)=-\gamma /|q| $
in $ d=1 $ has nothing to do with a frequently encountered artifact
when discretizing the functional integral inappropriately---sometimes
referred to as ``path collapse''.\cite{Kleinert}
This artifact occurs even in $ d \ge 2 $.
%
%
\section{Functional--analytic approach}

In this section we present the main results of the functional--analytic
approach to the one--dimensional hydrogen atom. The strategy
we are following is the standard one\cite{Rellich,AkhiezerG,ReedSII}
and was used to treat related problems or special
cases.\cite{Rellich,BullaG,AlbeverioG,CarreauF90a,AlbeverioB}
The aim is to find a dense domain in the Hilbert space
$ L^{2}({\Bbb{R}}) $, on which the formal differential operator
$ (-1/2)\d^{2}/\d q^{2} - \gamma /|q| $ is self--adjoint.

As a first step
it is often reasonable to define a formally given operator on the space
$ {\cal C}_{0}^{\infty } ({\Bbb{R}} ) $ of arbitrarily often differentiable
complex--valued functions with compact support in $ {\Bbb{R}} $. This is
not possible here because of the $ 1/|q| $--singularity. Therefore we
start with the operator $ h:= (-1/2) \d^{2}/\d q^{2} - \gamma /|q| $
on the domain
$ {\cal D}(h) := {\cal C}_{0}^{\infty } ({\Bbb{R}} \backslash \{ 0 \} ) $.
The operator $ h $ is symmetric, hence closable.
The operator closure $ \bar{h} $ of $ h $ has deficiency
indices $ (2,2) $, as can be seen by
solving explicitly the corresponding eigenvalue problem in
terms of Whittaker's functions.\cite{Kratz} It follows\cite{AkhiezerGPar127}
that there is
a four--parameter family $ \{H_{\bf B}\} $ of self--adjoint extensions of
$ \bar{h} $ parametrizable by the self--adjoint $ 2 \times 2 $--matrix
$ {\bf B} = \left( \begin{array}{cc}  \rho  & \zeta\\ \zeta^{*}
& \lambda  \end{array}
\right) \,,\;\, \lambda ,\rho   \in {\Bbb{R}} \; , \; \zeta \in {\Bbb{C}}
$,
\begin{equation}
H_{\bf B} := -\frac{1}{2}\;\frac{\d ^{2}}{ \d q^{2}} - \frac{\gamma }{|q|}
\end{equation}
\begin{eqnarray}
{\cal D}(H_{\bf B}):= \Bigg\{\varphi \in L^{2}({\Bbb{R}}) & :
& \varphi,\varphi ' \in
{\cal AC}_{\rm loc} ({\Bbb{R}} \backslash \{ 0 \}),
\left(-\frac{1}{2}\; \varphi '' - \frac{\gamma }{|q|}
\varphi \right)\in L^{2}({\Bbb{R}}), \nonumber\\[2mm]
& & \left( \begin{array}{l}
\displaystyle{\lim_{q \downarrow 0 }}\; \{ 2\gamma \varphi (q)
\ln (|\gamma|  q) + \varphi '(q) \} \\
\displaystyle{\lim_{q \uparrow 0 }}\; \{ 2\gamma \varphi (q)
\ln (-|\gamma| q) - \varphi '(q) \} \end{array} \right)
= {\bf B}
\left( \begin{array}{c} \displaystyle{\lim_{q \downarrow 0 }}\; \varphi (q) \\
\displaystyle{\lim_{q \uparrow 0 }}\; \varphi (q) \end{array} \right)
\Bigg\}\; .
\end{eqnarray}
Here $ {\cal AC}_{\rm loc} (\Lambda  ) $ denotes the set of complex--valued
functions $ \varphi $ on $ \Lambda \subseteq {\Bbb{R}} $
which are absolutely continuous on every compact subset of $ \Lambda $,
and the first and second derivatives $ \varphi ' $ and $ \varphi '' $
of $ \varphi $ are understood to be defined Lebesgue almost everywhere.
As limiting cases, we also
allow the entries of the matrix $ {\bf B} $ to be infinite in order not to
have to distinguish too many different cases. For example,
one way to arrive at the Dirichlet condition
$ \lim_{q \uparrow 0 }\; \varphi (q)= 0 =
\lim_{q\downarrow 0 }\;\varphi (q) $ at the origin is by choosing
$ \zeta=0 $, $ \lambda =\rho =\infty $.
We denote the corresponding Hamiltonian by $ H_{\rm D} $.
In the zero--coupling case $ \gamma =0 $, each $ H_{\bf B} $ describes
a particle on the real line subjected to a certain point
interaction at the origin.\cite{CarreauF90a,AlbeverioB}
In particular, for appropriate $ {\bf B} $ the $ \delta $-- and
(so--called) $ \delta ' $--interaction emerge.

Remarkably, there is a closed--form expression for the
energy--dependent Green function
$ \langle q_{b}|\left( H_{\bf B}-E\right)^{-1}|q_{a} \rangle $,
that is, for the integral kernel of the resolvent of $ H_{\bf B} $.
Since this expression involves the Green function of $ H_{\rm D} $, we
first state
\begin{eqnarray}\label{result}
\langle q_{b}|\left( H_{\rm D}-E\right)^{-1}|q_{a} \rangle
& = &
\Theta (q_{b}q_{a})\;
\frac{\Gamma (1-\gamma /\sqrt{-2E} )}{\sqrt{-2E}}    \nonumber\\
&&\times \bigg[\Theta (|q_{b}|-|q_{a}|) \;
W_{\gamma /\sqrt{-2E} ,1/2}(\sqrt{-8E} |q_{b}|) \;
M_{\gamma /\sqrt{-2E} ,1/2}(\sqrt{-8E}| q_{a}|) \nonumber\\
&& \hspace*{7mm} + ~( q_{b} \longleftrightarrow q_{a} ) \bigg]\; .
\end{eqnarray}
Here $ \Theta $ denotes Heaviside's unit--step function,
$ \Gamma $ denotes Euler's gamma function  and
$ M_{\mu , \tau } $ and $ W_{\mu , \tau } $ denote
Whittaker's functions.\cite{Kratz}
For $ q_{a}, q_{b} \ge 0 $, Eq.\ (\ref{result}) is just the Green
function of the hydrogen atom on the positive half--line with a Dirichlet
condition at the origin. Being the {\em s}--wave part of the
Green function of the three--dimensional hydrogen atom (apart from a factor
$ 4\pi q_{b}q_{a} $), it appeared for the first time in the early work
of Meixner\cite{Meixner} as a by--product. The second part of the next
section will be devoted to a genuine functional--integral derivation of
(\ref{result}).

For completeness and convenience of the reader we recall for the
attractive case $ \gamma > 0 $ the twofold
degenerate discrete eigenvalues $ E_{n}^{\rm (D)} $ and give an orthonormal
basis $ \{ \varphi _{n,1}^{\rm (D)}, \varphi _{n,2}^{\rm (D)} \} $ of the
corresponding two--dimensional eigenspaces of $ H_{\rm D} $ as obtained
from the poles and corresponding residues of~(\ref{result})
\begin{equation}\label{dirichletenergy}
E_{n}^{\rm (D)} = -\,\frac{\gamma ^{2}}{2n^{2}}\;,\qquad\quad n=1,2,3,\ldots\;,
\end{equation}
\begin{equation}
\varphi _{n,k}^{\rm (D)} (q)= \Theta \Big((-1)^{k}q\Big)
\frac{2\gamma^{3/2}}{n^{5/2}}\; |q| \; \e^{- |q| \gamma /n}\;
L_{n-1}^{(1)}(2 \, |q| \, \gamma /n) \;,\qquad\quad k=1,2\;.
\end{equation}
Here $ L_{n}^{(\alpha )} $ denotes a (generalized) Laguerre
polynomial.\cite{Kratz}

Green functions of different self--adjoint extensions of a closed
symmetric operator are related via a Krein formula.\cite{AlbeverioGAppA}
Its particularization to the case of the one--dimensional hydrogen atom
gives
\begin{eqnarray}\label{greenHB}
\langle q_{b}|\left( H_{\bf B}-E\right)^{-1}|q_{a} \rangle
& = &
\langle q_{b}|\left( H_{\rm D}-E\right)^{-1}|q_{a} \rangle
 + 2 \Big( \Gamma (1-\gamma /\sqrt{-2E} ) \Big) ^{2}
\nonumber\\
& & \hspace*{1cm}\times W_{\gamma /\sqrt{-2E}, 1/2}(\sqrt{-8E}\,|q_{b}|)
W_{\gamma /\sqrt{-2E}, 1/2}(\sqrt{-8E}\,|q_{a}|)
\nonumber\\
& & \hspace*{1cm}\times (\Theta (q_{b}) , \Theta (-q_{b}))
\Big( {\bf B} - \varepsilon _{\gamma }(E) \openone \Big)^{-1}
\left( \begin{array}{c} \Theta (q_{a}) \\
\Theta (-q_{a}) \end{array} \right) \,,
\end{eqnarray}
where
\begin{equation}\label{HL}
\varepsilon _{\gamma }(E) :=
2\gamma  \left[\ln(|\gamma |/\sqrt{-8E})-
\psi (1-\gamma /\sqrt{-2E}) + 2\psi (1) \right] -\sqrt{-2E}
\end{equation}
and $ \psi (z) := \d \ln \Gamma (z)/\d z $ is the
logarithmic derivative of Euler's gamma function.

Since the deficiency indices of $ \bar{h} $ are finite, all
self--adjoint extensions $ H_{\bf B} $ possess\cite{AkhiezerGSeite365}
the same essential
spectrum $ \sigma _{\rm ess}(H_{\bf B})=[0,\infty [ $.
For the same reason and due to the fact that $ H_{\rm D} $ is
bounded below, see e.g.\ Eq.\ (\ref{dirichletenergy}) above, one
concludes\cite{ReedSIIPropositionSeite179} that they are all bounded
below. The discrete spectrum of $ H_{\bf B} $ is obtained from the poles
of (\ref{greenHB}) and consists of those
$ E\in \; ]-\infty ,0[ $ for which $ \varepsilon _{\gamma }(E) $ is an
eigenvalue of $ {\bf B} $
\begin{equation}
2 \varepsilon _{\gamma }(E)=\lambda  +\rho \; \pm
\sqrt{ ( \lambda -\rho )^{2} + 4|\zeta|^{2} }\;.
\end{equation}
A plot of $ \varepsilon _{\gamma } /\gamma $ for $ E<0 $
as a function of $ \gamma /\sqrt{-2E} $ is given in Figure 1.
It turns out that the discrete spectrum is
twofold degenerate, if $ \zeta=0 $ and $ \lambda =\rho $.
Not unexpectedly, for $ \gamma  > 0 $ there are infinitely many discrete
eigenvalues irrespective of the choice of $ {\bf B} $. Even for
$ \gamma  < 0 $ there may be one or two negative discrete eigenvalue(s).
For a given coupling constant $ \gamma \in {\Bbb{R}} $ one can choose the
matrix $ \bf B $ and
thus the boundary condition in such a way that the ground--state energy
of $ H_{\bf B} $ is arbitrarily negative.

An interesting quantity to look at is the probability
current density $ j_{\varphi }(q) := {\rm Im} \{\varphi ^{*}(q)
\varphi ' (q) \} $ for $ \varphi \in {\cal D}(H_{\bf B}) $.
Due to self--adjointness it is continuous in $ q $ for all
$ q\in{\Bbb{R}} $, even at the origin, where it is given by
\begin{equation}
j_{\varphi }(0)= {\rm Im} \{ \zeta \,\lim\limits_{q\downarrow 0}
\varphi ^{*}(q)\varphi (-q)\}\,.
\end{equation}
In consequence, the choice $ \zeta =0 $ implies
$ j_{\varphi }(0)=0 $ for all $ \varphi \in {\cal D}(H_{\bf B}) $, which means
that the real line is effectively split into its
negative and positive half by the $ 1/|q| $--singularity.
It is only the resulting two--parameter subfamily that
can be obtained as the orthogonal sum of two Hamiltonians each of which
is defined
separately on the Hilbert space corresponding to one of the half--lines.
Thorough discussions of the one--parameter family of
Hamiltonians associated with
the one--dimensional hydrogen atom on the
half--line have been given earlier,
see Refs.\ \onlinecite{Rellich,BullaG,AlbeverioG}.
%
%
\section{Functional--integral approach}

As already mentioned in the Introduction, the Wiener integral on the
{\sc rhs} of (\ref{feynmankac}) for the
$d$--dimensional hydrogen atom is well--defined and finite for all
$ d \ge 2 $. By now, also the explicit
evaluation\cite{DuruK,BlanchardS,HoI,Inomata,Steiner84,PakS,Steiner,%
ChetouaniH,YoungD,Kleinert87,CastrigianoS89,CastrigianoS90,CastrigianoS91}
of the Laplace transform of such functional integrals is well--established.
The one--dimensional case was explicitly
considered in Ref.\onlinecite{ChetouaniH} by formally setting $ d=1 $
in the final result of such an evaluation without any justification.
The trouble with this procedure is that one does
not know {\em a priori} to which physical problem it yields the solution.
Instead, as a first step it is necessary to discuss the question of how the
hydrogen atom can be reasonably defined in one dimension from the point
of view of functional integration.

To begin with, we clarify for which paths $ x $ of the one--dimensional
Wiener process with starting point $ q_{a}\in {\Bbb{R}} $ the integral
$ \int_{0}^{t}\d t'/|x(t')| $ is finite.
Due to continuity, one has the following inclusion
between path sets
\begin{eqnarray}\label{incl1}
{\cal N}_{t}({\Bbb{R}} , q_{a}) & := & \Big\{ x \in {\cal C}({\Bbb{R}},q_{a}):
x(t')\neq 0 \quad\mbox{for all}\quad
t'\in [0,t]\Big\}\nonumber\\
& \subseteq &
\Big\{x\in {\cal C}({\Bbb{R}},q_{a}):
\int\limits_{0}^{t}\frac{\d t'}{|x(t')|} < \infty \Big\}\;.
\end{eqnarray}
Introducing the local time\cite{RevuzY} of $ x $ at $ q $, which is
formally given by $ L_{t,q}[x] = \int_{0}^{t}\d t'\,
\delta ^{(1)}(x(t')-q) $, one can write
\begin{equation}
\int\limits_{0}^{t}\frac{\d t'}{|x(t')|} =
\int\limits_{{\Bbb{R}}}\frac{\d q}{|q|}\;L_{t,q}[x]\;.
\end{equation}
Hence we deduce a further inclusion
\begin{equation}\label{incl2}
\Big\{x\in {\cal C}({\Bbb{R}},q_{a}):
\int\limits_{0}^{t}\frac{\d t'}{|x(t')|} < \infty \Big\}
\;\subseteq\; \Big\{ x\in {\cal C}({\Bbb{R}},q_{a}): L_{t,0}[x] =0\Big\}\;.
\end{equation}
According to Ref.\ \onlinecite{RevuzYRemark}, the set
$ {\cal N}_{t}({\Bbb{R}},q_{a}) $
equals almost surely (a.s.) the set on the {\sc rhs} of (\ref{incl2}).
Therefore one infers the a.s.--equality
\begin{equation}\label{gleich}
{\cal N}_{t}({\Bbb{R}},q_{a}) \;=\;
\Big\{x\in{\cal C}({\Bbb{R}},q_{a}):
\int\limits_{0}^{t}\frac{\d t'}{|x(t')|} < \infty \Big\}\;,
\end{equation}
which is a special case of the Theorem on p. 785 in Ref.\ \onlinecite{EzawaK}.

After this prelude the following definition for the functional integral of
the one--dimensional hydrogen atom will suggest itself
\begin{equation}\label{1hpi}
\int\limits_{{\cal N}_{t}({\Bbb{R}},q_{a})}
\d W^{1}[x]\;\delta^{(1)} (x(t)-q_{b})\;
\exp \left\{\int\limits_{0}^{t}\d t'\;\frac{\gamma }{|x(t')|}\right\} \; .
\end{equation}
Five remarks are in order.
\begin{itemize}
\item[ (i)]
Cor.\ 6.3.10 in Ref. \onlinecite{BratteliR} implies that the
restriction of the Wiener measure to $ {\cal N}_{t}({\Bbb{R}},q_{a}) $
yields again a
Markovian path measure which is invariant under time reversal, namely the
measure corresponding to the Wiener process on
$ {\Bbb{R}} $ killed at the origin. Using this fact and
\begin{equation}
\lim_{t \downarrow 0} \;\sup_{q\in{\Bbb{R}}}
\int\limits_{{\cal N}_{t}({\Bbb{R}},q)} \d W^{1}[x]\;
\int\limits_{0}^{t}\frac{\d t'}{|x(t')|} = 0 \; ,
\end{equation}
it follows from the appropriate versions of
Nakao's inequality\cite{Nakao} and
Khas'minskii's lemma\cite{AizenmanS} that the
functional integral (\ref{1hpi}) exists.
\item[ (ii)]
Similar to the proof of Thm.\ B.1.1 in Ref.\ \onlinecite{Simon82},
the Cauchy--Schwarz inequality together with (i) imply that (\ref{1hpi})
is the integral kernel of a self--adjoint semigroup on $ L^{2}({\Bbb{R}}) $.
This semigroup is strongly continuous, as can be seen by adapting the
proof of Thm.\ 3.1 in Ref.\ \onlinecite{BroderixH}, see also
Prop.\ 3.2 in Ref.\ \onlinecite{Carmona79}.
\item[ (iii)]
The generator of the semigroup defined by
(\ref{1hpi}) is the Dirichlet Hamiltonian $ H_{\rm D} $ of Sec.~II.
This follows from the Feynman--Kac
formula of Thm.\ 6.3.12 in Ref.\ \onlinecite{BratteliR},
and the fact that $ 1/|q| $ is operator bounded with respect
to the Dirichlet Laplacian with relative bound zero.\cite{Gesztesy}
\item[ (iv)]
Restricting the paths to the set $ {\cal N}_{t}({\Bbb{R}},q_{a}) $
is necessary for
(\ref{1hpi})  to exist in the attractive case $ \gamma > 0 $.
\item[ (v)]
In the repulsive case $ \gamma < 0 $ the restriction does not
have any effects,
if one adopts the convention $ \e^{-\infty } := 0 $.
This illustrates the fact that the existence of the unrestricted
Wiener integral on the
{\sc rhs} of (\ref{feynmankac}) does not necessarily imply that the
functional--analytic approach leads to a unique Hamiltonian.
\end{itemize}

In the second part of this section we will sketch how to obtain the Green
function of $ H_{\rm D} $ by computing
\begin{eqnarray}\label{green}
\langle q_{b}|\left( H_{\rm D}-E\right)^{-1}|q_{a} \rangle & = &
\int\limits_{0}^{\infty }\d t\;\e^{Et}\;
\langle q_{b}|\e^{- t H_{\rm D}}|q_{a} \rangle \nonumber\\
& = & \int\limits_{0}^{\infty }\d t\;\e^{Et}
\int\limits_{{\cal N}_{t}({\Bbb{R}}, q_{a})}
\d W^{(1)}[x]\;\delta^{(1)} (x(t)-q_{b})\;
\exp \left\{\int\limits_{0}^{t}\d t' \frac{\gamma }{|x(t')|}\right\}
\end{eqnarray}
for complex energies $ E $ with a real part smaller than the infimum of the
spectrum of $ H_{\rm D} $. Thereby we follow ideas presented in
Refs. \onlinecite{DuruK,BlanchardS,HoI,Inomata,Steiner84,PakS,Steiner,%
ChetouaniH,YoungD,Kleinert87,CastrigianoS89,CastrigianoS90,%
CastrigianoS91,FLM1,FLM2,FLM3},
where a more recent technique in functional integration, namely
path--dependent time transformations, has been developped for
quantum--mechanical applications.
It is only with the help of
this technique that a number of potential systems have been treated
exactly in the framework of functional integration.
We are going to present only its main ideas,
but all steps which are needed for the calculation of (\ref{green}) will
be explained. For a detailed and fairly pedagogical approach to
path--dependent time
transformations, the reader may consult
Refs.\
\onlinecite{YoungD,CastrigianoS89,CastrigianoS90,CastrigianoS91,FLM1}.

In what follows it will be convenient to introduce the Bessel
process\cite{ItoM} with index $ 1/2\, $, which is a continuous
Markov process taking
on values in the positive half--line $ {\Bbb{R}}_{+} $.
It is characterized by the transition density
\begin{equation}
b_{t}^{(1/2)}(q',q) := \frac{q'}{q}\,(2\pi t)^{-1/2}\,
\left( \exp\left\{ -\frac{(q'-q)^{2}}{2t}\right\}
- \exp\left\{ -\frac{(q'+q)^{2}}{2t}\right\}\right)\;,\quad
q',q\in{\Bbb{R}}_{+}\;.
\end{equation}
Hence, $ \Theta(qq')(q/q') b_{t}^{(1/2)}(|q'|,|q|)\;$, $ q',q\in{\Bbb{R}} $, is
the (non--normalized) transition density of the Wiener process
on $ {\Bbb{R}} $ killed
at the origin and the
second line in (\ref{green}) can be rewritten\cite{eigenlob} as
$ \Theta(q_{b}q_{a})q_{a}/q_{b} $ times
\begin{equation}\label{star}
\int\limits_{0}^{\infty }\d t\;\e^{Et}
\int\limits_{{\cal C}({\Bbb{R}}_{+}, |q_{a}|)}
\d B_{1/2}[x]\;\delta^{(1)} (x(t)-|q_{b}|)\;
\exp \left\{\int\limits_{0}^{t}\d t' \frac{\gamma }{x(t')}\right\}\; .
\end{equation}
Here $ \d B_{1/2} $ denotes the probability measure associated with the Bessel
process with index $ 1/2 $.

The key for the calculation of (\ref{star}) is the equality of measures
\begin{equation}\label{measure}
\d B_{1/2} = {\sf S}(\d W^{4})\;.
\end{equation}
It states that the measure of the Bessel process with index $ 1/2 $
arises as the image of the four--dimensional Wiener measure under the
transformation $ {\sf S} $, which maps the set
$ {\cal C}({\Bbb{R}}^{4}, r_{a}) $ of paths $ y $ onto the set
$ {\cal C}({\Bbb{R}}_{+}, |q_{a}|) $ of paths $ x $ according to
\begin{equation}\label{cov}
x(t)= ({\sf S}y)(t) := [{ y}(s_{{ y}}(t))]^{2} \; , \qquad
{ r}_{a}^{2}:=|q_{a}|\; .
\end{equation}
Here $ s_{{ y}}(t) $ is defined to be
the inverse mapping of the path--dependent time
\begin{equation}
t_{{ y}}(s):=\int\limits_{0}^{s}\d s'\; 4\,({ y}(s'))^{2}\;,
\end{equation}
that is, $ t_{{ y}}(s_{{ y}}(t))=t $.
Taking (\ref{measure}) for granted, for the time being,
we can rewrite (\ref{star}) in terms of the paths $ y $
\begin{equation}\label{massgl}
\mbox{(\ref{star})} = 4|q_{b}|\int\limits_{0}^{\infty }\d s\;\e^{4\gamma s}
\int\limits_{{\cal C}({\Bbb{R}}^{4},{ r}_{a})}
\d W^{4}[{ y}]\;\delta^{(1)} \Big(({ y}(s))^{2} - |q_{b}|\Big)\;
\exp \left\{4E\int\limits_{0}^{s}\d s'\; ({ y}(s'))^{2}\right\} \;.
\end{equation}
In doing so, we have performed the change--of--variables
$ t'\mapsto s':=s_{{ y}}(t') $, which is possible since
$ t_{y}(s) $ is\cite{2.7RevuzY} $ \d W^{4} $--a.s.\ strictly increasing
in $ s $,
and we have used $ t_{y}(\infty)=\infty $
$ ~~\d W^{4} $--a.s.\cite{7.12Simon}

In order to calculate (\ref{massgl}) we remark that
\begin{equation}
\delta ^{(1)}\Big(({ y}(s))^{2} - |q_{b}|\Big) =
\frac{|q_{b}|}{2}
\int\d^{3}\Omega _{{ r}_{b}}\;\delta ^{(4)}({ y}(s)-{ r}_{b})\; ,
\end{equation}
where $ \int\d ^{3}\Omega _{{ r}_{b}} \; (\bullet) $
stands for the integration with respect to the usual rotational invariant
measure on the unit sphere in $ {\Bbb{R}} ^{4} $ over the
three angles of $ r_{b} \in {\Bbb{R}} ^{4} $ in spherical co--ordinates.
Using the well--known explicit expression for the integral kernel of the
semigroup generated by the
four--dimensional isotropic harmonic oscillator\cite{Simon79,Roepstorff}
with spring constant $ \omega ^{2}:= -8E $,
the {\sc rhs} of (\ref{massgl}) can be cast into the form
\begin{eqnarray}
\lefteqn{\frac{q_{b}^{2}}{2}\,\left(\frac{\omega }{\pi }\right)^{2}
\int\limits_{0}^{\infty }\d s\; \frac{\exp\left\{ 4\gamma s\right\}}
{[\sinh (\omega s)]^{2}} \; \exp\left\{-\frac{\omega }{2}\;
(|q_{b}|+|q_{a}|)\; \coth (\omega s)\right\}}\nonumber\\
&&
\hspace*{4.5cm}\times\int\d^{3}\Omega _{{ r}_{b}}\; \exp\left\{\frac{\omega }
{\sinh (\omega s)}\; { r}_{b}\cdot{ r}_{a}\right\}\;.
\end{eqnarray}
Applying the formula\cite{Erdelyi}
\begin{equation}
\int\d^{3}\Omega _{{ r}_{b}}\; \exp\left\{z\,
{ r}_{b}\cdot{ r}_{a}\right\} =
\frac{(2\pi )^{2}}{z|{ r}_{b}||{ r}_{a}|}\;
I_{1}(z|{ r}_{b}||{ r}_{a}|)
\end{equation}
($ z\in{\Bbb{R}}$, $ I_{\nu } $ denotes the modified Bessel function of
the first kind\cite{Kratz} with index $ \nu $) and performing the Laplace
transformation,\cite{6.669.4} the expression (\ref{star}) is eventually
seen to be identical to
\begin{equation} \label{final}
(2|q_{b}|/\omega |q_{a}|)\;\Gamma (1-2\gamma /\omega) \;
\Big( \Theta (|q_{b}|-|q_{a}|) \; W_{2\gamma /\omega ,1/2}(\omega |q_{b}|) \;
M_{2\gamma /\omega ,1/2 }(\omega |q_{a}|) \;
+ \;
( q_{b} \longleftrightarrow q_{a} ) \Big).
\end{equation}
Since the Green function is analytic\cite{ReedSI}
on the complement of the spectrum
of $ H_{\rm D} $, the desired result (\ref{result}) follows from
(\ref{final}), provided that (\ref{measure}) holds.

Following Ref.\ \onlinecite{FLM1}, the validity of the equality of
measures (\ref{measure})  with $ {\sf S} $ as in
(\ref{cov}) is implied by the equality
\begin{eqnarray}\label{freiegl}
\lefteqn{\int\limits_{0}^{\infty }\d t\;\e^{Et}\;
\int\limits_{{\cal C}({\Bbb{R}}_{+}, |q_{a}|)}
\d B_{1/2}[x]\;\delta^{(1)} (x(t)-|q_{b}|)}\nonumber\\
&& \hspace*{1.2cm}= 4|q_{b}|\int\limits_{0}^{\infty }\d s
\int\limits_{{\cal C}({\Bbb{R}}^{4},{ r}_{a})}
\d W^{4}[{ y}]\;\delta^{(1)} \Big(({ y}(s))^{2} - |q_{b}|\Big)\;
\exp \left\{4E\int\limits_{0}^{s}\d s'\; ({ y}(s'))^{2}\right\} \;,
\end{eqnarray}
which is nothing but the free--particle version $ \gamma =0 $ of
(\ref{massgl}).

To verify (\ref{freiegl}) we reduce (\ref{final}) for the case
$ \gamma =0 $ with the help of Ref.\ \onlinecite{MagnusO} to
\begin{equation}
\frac{|q_{b}/q_{a}|}{\sqrt{-2E}}\;\Big(\exp\left\{\sqrt{-2E}\,
\big| |q_{b}|-|q_{a}|\big|\right\} -
\exp\left\{\sqrt{-2E}\, \big( |q_{b}|+|q_{a}|\big)\right\}\Big) \; ,
\end{equation}
which is indeed the Laplace transform\cite{3.471.9} of the transition
density $ b_{t}^{(1/2)}(|q_{b}|,|q_{a}|) $.

Within the theory of stochastic differential equations
there are other ways\cite{YoungD,CastrigianoS90,CastrigianoS91}
to prove the equality
of measures (\ref{measure}). In essence, they rely on
It\^{o}'s formula and on a theorem due to Dambis, Dubins and
Schwarz.\cite{DDSRevuzY}

To summarize, the path--dependent time transformation (\ref{cov}) relates
the energy--dependent Green function of the one--dimensional hydrogen
atom, which is the Laplace transform of the non--Gaussian functional integral
(\ref{1hpi}), to the Green function of a four--dimensional harmonic oscillator
(\ref{massgl}), that is, to the Laplace transform of a Gaussian
functional integral, which is explicitly known.
The r\^{o}le of the spring constant
is thereby played by the energy of the hydrogen atom and the r\^{o}le of
the energy of the harmonic oscillator is played by the
coupling constant $ \gamma $.

A path--dependent time transformation similar to (\ref{cov}) was used in
previous functional--integral calculations for the radial Green function of the
hydrogen atom in two or more dimensions.\cite{Inomata,Steiner84,Steiner,%
ChetouaniH,CastrigianoS90}
Further aspects of the relation between the hydrogen atom and the harmonic
oscillator in the framework of the stationary Schr\"odinger
equation may be found, e.g., in Refs.\ \onlinecite{Junker,BatemanB}.
%
%
\section{Discussion}

Clearly, mathematics alone cannot tell which
particular element of the four--parameter family of Hamiltonians
$ \{ H_{\bf B}\} $ should be chosen to model a
given experimental situation. Instead, additional physical information
is necessary. In this context it is useful to know that, for example, the
sequence of Hamiltonians corresponding to the cut--off potentials
$ -\gamma /(|q| +a) $ converges in uniform resolvent
sense to the Dirichlet Hamiltonian $ H_{\rm D} $ as
$ a \downarrow 0 $.\cite{Gesztesy}
Another fact which makes $ H_{\rm D} $ stand out is that the domains of all
other self--adjoint extensions contain wave functions for which the
expectation values
of both the kinetic and the potential energy are infinite---though their sum,
the total energy, is always finite.
Among the domains $ {\cal D}(H_{\bf B}) $ it is only
$ {\cal D}(H_{\rm D}) $ on which one
can define the kinetic energy $ -(1/2) \d^{2}/\d q^{2} $
as a self--adjoint operator. In this light, it is no
surprise that the functional--integral approach in Sec.~III is only able
to yield $ H_{\rm D} $. Indeed, there it was required to average
the Feynman--Kac functional
$ \exp\{ -\int_{0}^{t}\d t'\, V(x(t'))\} $ with respect to a path
measure generated by the kinetic energy. By abandoning this requirement, it is
shown in Refs.\ \onlinecite{EzawaK,FarhiG,CarreauF90b} that the
functional--integral approach can also provide self--adjoint extensions other
than the Dirichlet Hamiltonian. The price one has to pay is a suitable
regularization of the potential near the singularity and a subsequent limit.
For an interesting construction of a family of path measures corresponding to
$ \{ H_{\bf B} \} $  in the zero--coupling case, see
Ref.\ \onlinecite{CarreauF90a}.

So far, there is no systematic functional--integral analogue of the calculus of
deficiency indices in the functional--analytic approach. To develop such an
analogue would be an interesting programme for future research.
%
%
\section*{Acknowledgments}

We are grateful to Kurt Broderix for his interest and for hints
concerning the functional--analytic treatment.
W.F.\ acknowledges support by the
Evangelisches Studienwerk Villigst (Schwer\-te, Germany).
This work was partially supported by the Human Capital and Mobility
Programme ``Polarons, bi--polarons and excitons. Properties and
occurrence in new materials.'' of the European Community.


%
%
%
%
%
\begin{figure}
\vspace{2cm}
\caption{Plot of $\varepsilon_{\gamma}/\gamma $ for $  E<0 $ as a function
  of $ \gamma /\protect{\sqrt{-2E}} $, where $ \varepsilon _{\gamma } $
  is defined in Eq. (\protect{\ref{HL}}). The value of
  $\varepsilon_{\gamma}/\gamma $ approaches $ 4 \psi (1) - 2
  \ln 2 = -3.695\ldots $ as $ \gamma /\protect{\sqrt{-2E}} \to - \infty $.}
\end{figure}
\end{document}